\documentclass[aps,pra,twocolumn]{revtex4-2}
\usepackage{color}
\usepackage{bm}
\usepackage{braket}
\usepackage{nccmath}
\usepackage{upgreek}
\usepackage{graphicx}
\usepackage{makecell}
\usepackage{hyperref}
\hypersetup{colorlinks=true, citecolor=blue, urlcolor=blue, linkcolor=blue}

\begin{document}
\title{Optimal Selective Orientation of Chiral Molecules Using Femtosecond Laser Pulses} 
\author{Long Xu}
\affiliation{AMOS and Department of Chemical and Biological Physics, The Weizmann Institute of Science, Rehovot 7610001, Israel}
\author{Ilia Tutunnikov}
\affiliation{AMOS and Department of Chemical and Biological Physics, The Weizmann Institute of Science, Rehovot 7610001, Israel}
\author{Yehiam Prior}
\email{yehiam.prior@weizmann.ac.il}
\affiliation{AMOS and Department of Chemical and Biological Physics, The Weizmann Institute of Science, Rehovot 7610001, Israel}
\author{Ilya Sh. Averbukh}
\email{ilya.averbukh@weizmann.ac.il}
\affiliation{AMOS and Department of Chemical and Biological Physics, The Weizmann Institute of Science, Rehovot 7610001, Israel}

\begin{abstract}
We present a comprehensive study of enantioselective orientation of chiral molecules excited by a pair of delayed cross-polarized femtosecond laser pulses.
We show that by optimizing the pulses' parameters, a significant ($\sim 10\%$) degree of enantioselective orientation can be achieved at zero and at five kelvin rotational temperatures. This study suggests a set of reasonable experimental conditions for inducing and measuring strong enantioselective orientation. The strong enantioselective orientation and the wide availability of the femtosecond laser systems required for the proposed experiments may open new avenues for discriminating and separating molecular enantiomers.

\end{abstract}
\maketitle

\section{Introduction}
Chiral molecules exist in two mirror-symmetric forms: left- and right-handed enantiomers \cite{Cotton1990Chemical}.
Ever since the discovery of molecular chirality by Louis Pasteur in the 19th century \cite{Pasteur1848}, the separation of enantiomers has attracted significant attention due to the enantioselectivity of various chemical, physical, and biological processes. However, since many of the enantiomers' physical properties, such as melting point and mass, are identical, enantiomer differentiation is a very challenging and elusive task.
Several approaches to chiral discrimination which are based on electromagnetic fields have been put forward and demonstrated in recent decades. Among others, these include photoelectron circular dichroism \cite{Ritchie1976Theory,Bowering2001Asymmetry,Lux2011Circular,Beaulieu2017Attosecond,Beaulieu2018Photoexcitation}, Coulomb explosion imaging \cite{Pitzer2013Direct,Herwig2013Imaging,Fehre2019Enantioselective}, microwave three-wave mixing spectroscopy \cite{patterson2013enantiomer,Patterson2013Sensitive,Patterson2014New,Alvin2014Enantiomer,lehmann2018theory,Ye2018Real,Ye2019Determination,leibscher2019principles},  high-order harmonic generation \cite{Cireasa2015Probing,Ayuso2021}, and enantiospecific molecular interaction with achiral magnetic substrates \cite{Banerjee-Ghosh2018}.

Over the last several years, enantioselective rotational control of chiral molecules in the gas phase using pulsed electromagnetic fields has been studied theoretically. The list of studies includes enantioselective orientation by laser pulses with twisted polarization \cite{Yachmenev2016Detecting,Gershnabel2018Orienting,Tutunnikov2018Selective,Milner2019Controlled,Tutunnikov2019Laser,Tutunnikov2020Observation, xu2021enantioselective}, terahertz  pulses with twisted polarization \cite{Tutunnikov2021Enantioselective}, and cross-polarized two-color laser pulses \cite{Takemoto2008Fixing, Xu2021Three}. 
Selective chiral control was experimentally demonstrated in propylene oxide molecules excited by an optical centrifuge \cite{Milner2019Controlled}.
An optical centrifuge \cite{Karczmarek1999Optical,Villeneuve2000Forced,Yuan2011Dynamics,Korobenko2014Direct,Korobenko2018Control} is a unique and extreme example of a laser pulse with twisted polarization, where the laser polarization vector rotates unidirectionally in plane with increasing angular velocity.
Other examples that have been discussed include delayed cross-polarized laser pulses \cite{Fleischer2009Controlling,Kitano2009Ultrafast,Khodorkovsky2011Controlling}, polarization-shaped pulses \cite{Kida2008Stimulated,Kida2009Coherent,Karras2015Polarization,Prost2017Third,Mizuse2020Direct}, and chiral pulse trains \cite{Zhdanovich2011Control,Johannes2012Molecular}. While the optical centrifuge is a relatively efficient tool for inducing enantioselective orientation, achieving similar degrees of orientation using, e.g., delayed cross-polarized laser pulses will require a much higher peak power \cite{Tutunnikov2019Laser}, which may lead to a non-negligible molecular ionization.

In this work, we optimize the parameters of delayed cross-polarized femtosecond laser pulses for achieving a high degree of enantioselective orientation, while keeping the peak power low to minimize ionization. We consider propylene oxide as a typical example of a small chiral molecule. To the best of our knowledge, there were no comprehensive studies of the dependence of the enantioselective orientation on the pulses' parameters. We show that by correctly choosing the intensities of the two pulses and the relative delay, a high degree of enantioselective orientation can be achieved. As femtosecond laser pulses are extensively used and are widely available in a large number of laboratories, we hope that our results will stimulate further experimental activity towards laser-induced enantioselective orientation.

\section{Numerical methods}
We model the molecule as a rigid body and carry out quantum mechanical and classical simulations. The Hamiltonian for the laser-driven molecular rotation is given by $H(t)=H_{r}+H_{\mathrm{int}}(t)$ \citep{Krems2018Molecules,Koch2019Quantum}, where $H_{r}$ is the rotational kinetic energy of the  molecule and $H_{\mathrm{int}}(t)=-\mathbf{E}(t)\cdot[\bm{\alpha}\mathbf{E}(t)]/2$ is the field-molecule interaction. 
Here $\bm{\alpha}$ is the molecular polarizability tensor and $\mathbf{E}(t)$ is the electric field. 
The field-free symmetric-top eigenfunctions $|JMK\rangle$ are used as a basis set in our quantum-mechanical simulations \citep{zare1988Angular}.
Here $J=0,1,\dots$ is the total angular momentum, $-J\leq M\leq J$ and $-J\leq K \leq J$ are the angular momentum projections on the laboratory $Z$ and the molecular $z$ axes, respectively. 
The time-dependent Schr\"{o}dinger equation $i\hbar\partial_{t}|\Psi(t)\rangle=H(t)|\Psi(t)\rangle$ is solved by numerical exponentiation of the Hamiltonian matrix \citep{sidje1998Expokit}.

In the classical limit, we solve a coupled system combining Euler's equations and a quaternion's equation of motion.
In the frame of the principal axes of the inertia tensor, Euler's equations are expressed  as $\mathbf{I}\bm{\dot{\Omega}}=(\mathbf{I}\bm{\Omega})\times\bm{\Omega}+\mathbf{T}$ \citep{Goldstein2002Classical},
where $\bm{\Omega}=(\Omega_{x},\Omega_{y},\Omega_{z})$ is the angular velocity vector, $\mathbf{I}=\mathrm{diag}(I_{x},I_{y},I_{z})$ is the moment of inertia tensor, and $\mathbf{T}=(T_{x},T_{y},T_{z})$ is the external torque vector given by $\mathbf{T}=\bm{\alpha}\mathbf{E}(t)\times \mathbf{E}(t)$.
The relation between the laboratory and molecular frames is parametrized by a quaternion, $q=(q_{0},q_{1},q_{2},q_{3})$ \citep{Coutsias2004The,Kuipers1999Quaternions}. 
The quaternion's equation of motion is $\dot{q}=q\Omega/2$, with $\Omega=(0,\Omega_{x},\Omega_{y},\Omega_{z})$ \citep{Coutsias2004The,Kuipers1999Quaternions}.
We use $N=10^{6}$ molecules to simulate the behavior of a classical ensemble.
The molecules are initially isotropically distributed in space and their orientations are given by random uniform quaternions generated as described in \citep{Lavalle2006Planning}.
The initial angular velocities are given by the Boltzmann distribution $P(\Omega_i)\propto\exp\left(-I_{i}\Omega_{i}^{2}/2k_{B}T\right)$, $i=x,y,z$, where $T$ is the temperature and $k_{B}$ is the Boltzmann constant. A detailed description of our theoretical approaches can be found in \citep{Tutunnikov2019Laser}.  

In the present work, propylene oxide (PPO, $\mathrm{CH_3CHCH_2O}$) is used as an example of a typical chiral molecule. PPO molecules are excited by a pair of time-delayed cross-polarized laser pulses, where the first pulse is polarized along the $X$ direction and the polarization of the second pulse is in the $XY$ plane, at an angle $\pi/4$ with respect to the $X$ direction. 
The combined electric field is given by 
\begin{equation}
\mathbf{E}(t)=E_1  f(t)\cos(\omega t)\mathbf{e}_{X} + \frac{E_2}{\sqrt{2}}  f(t-\tau)\cos(\omega t)(\mathbf{e}_{X} + \mathbf{e}_{Y}).\label{eq:laser_Field}
\end{equation} 
Here the envelope is defined by $ f(t)=\exp\left(-2 \ln 2\,t^2/{\sigma^2}\right)$; $\omega$ is the carrier frequency; $E_1$ and $E_2$ are the peak amplitudes of the first and the second laser pulses; $\tau$ is the time delay between peaks of the two laser pulses and $\sigma$ is the full width at half maximum (FWHM) of the laser pulse intensity.
$\mathbf{e}_{X}$ and $\mathbf{e}_{Y}$ are unit vectors along the laboratory $X$ and $Y$ axes, respectively.

\begin{figure}[!b]
\centering{}
\includegraphics[width=\linewidth]{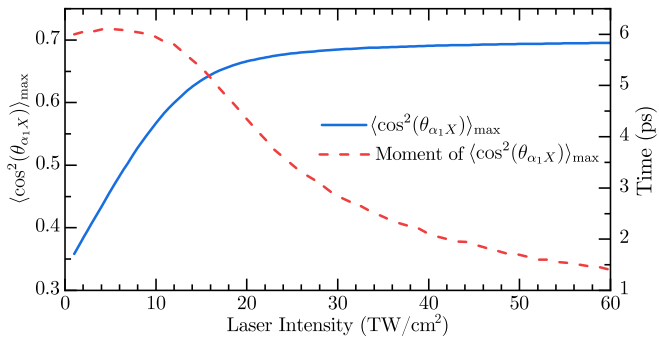}
\caption{Quantum mechanically calculated maximum degree of alignment, $\braket{\cos^2(\theta_{\alpha_1 X})}_\mathrm{max}$ and the time of its maximum as functions of the laser intensity. Here $T=0\,\mathrm{K}$ and $\sigma=50\, \mathrm{fs}$. 
 \label{fig:degree_alignment}}
\end{figure}

\section{Zero temperature case}
We start with the case of PPO molecules at zero rotational temperature $T=0\,\mathrm{K}$ (i.e., all molecules are in their ground rotational state, $\ket{0,0,0}$) excited by a single laser pulse.
Figure \ref{fig:degree_alignment} shows (solid blue line) the quantum mechanically calculated maximum degree of alignment, $\braket{\cos^2(\theta_{\alpha_1 X})}_\mathrm{max}$, as a function of the laser intensity. 
Here $\braket{\cos^2\,(\theta_{\alpha_1 X})}=\langle \Psi|(\bm{\alpha}_1\cdot \mathbf{e}_X)^2|\Psi\rangle$, where $\theta_{\alpha_1 X}$ denotes the angle between the most polarizable molecular axis, $\alpha_1$, and the laboratory-fixed $X$ axis (the polarization direction of the laser pulse).
The maximum degree of alignment, $\braket{\cos^2(\theta_{\alpha_1 X})}_\mathrm{max}$, increases monotonically with increasing laser intensity and tends to about 0.70 at the high-intensity limit.
This general behavior of the intensity-dependent degree of alignment is consistent with the previous works  \cite{Averbukh2004,Arnaud2008Towards}.
In addition, and as might be expected, the maximum alignment time decreases with increasing laser intensity (dashed red line).
When the molecules are excited by an increasingly intense laser pulse, additional excited rotational states are populated, resulting in a shorter response time.
Note that the degrees of alignment shown in this work are the same for the left- and right-handed enantiomers.

\begin{figure}[!b]
\centering{}
\includegraphics[width=\linewidth]{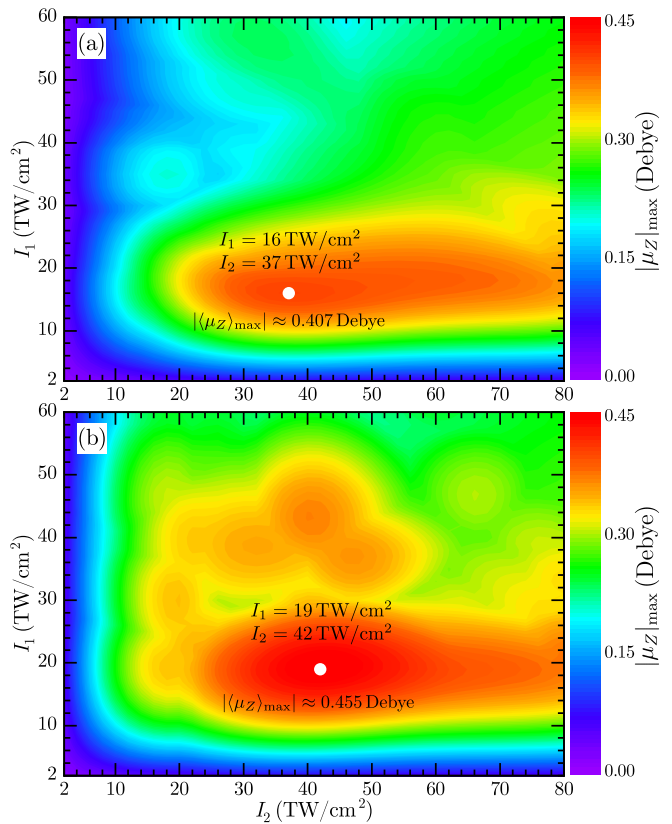}
\caption{Quantum mechanically calculated maximum (absolute value) of the dipole signal, $|\braket{\mu_Z}|_\mathrm{max}$, as a function of the peak intensities of the two laser pulses, $I_1$ and $I_2$.
Here $T=0\,\mathrm{K}$, $\sigma=50\, \mathrm{fs}$. The second laser pulse is applied (a) at the time of maximum degree of alignment induced by the first pulse,  $\braket{\cos^2(\theta_{\alpha_1 X})}_\mathrm{max}$ (see Fig. \ref{fig:degree_alignment}) and (b) at the optimal delay (producing the maximal degree of enantioselective orientation).
 \label{fig:figure1}}
\end{figure}

Next we consider excitation by a double pulse. A pair of properly delayed cross-polarized femtosecond laser pulses are known to be effective for inducing enantioselective orientation along the direction perpendicular to the plane spanned by the polarization vectors of the two pulses \cite{Yachmenev2016Detecting,Gershnabel2018Orienting,Tutunnikov2018Selective,Tutunnikov2019Laser}. Here, the enantioselective orientation is quantified by the expectation value of the molecular dipole projection along the propagation direction, $\braket{\mu_Z}=\langle \Psi|\bm{\mu}\cdot \mathbf{e}_Z|\Psi\rangle$, where $\bm{\mu}$ is the permanent dipole moment and $\mathbf{e}_Z$ is the unit vector along the laboratory $Z$ axis. Notice that the dipole signals of the two enantiomers have opposite signs but are equal in absolute value. We wish to find the optimal experimental conditions for enhanced enantioselective orientation. 

In Fig. \ref{fig:figure1}(a), we plot the maximum value of the transient dipole signal after the two-pulse excitation, $|\braket{\mu_Z}|_\mathrm{max}$ as a function of the laser intensities $I_1$ and $I_2$. The second cross-polarized laser pulse is applied when the degree of alignment, $\braket{\cos^2(\theta_{\alpha_1 X})}$, induced by the first laser pulse reaches its maximum value (see Fig. \ref{fig:degree_alignment}).
We search for the maximum value, $|\braket{\mu_Z}|_\mathrm{max}$, within the first 100\,ps after the excitations. 
The long-term field-free dipole signal may be affected by the centrifugal distortion and radiation emission due to rapid changes of the molecular dipole moment \cite{Babilotte2016Observation, Damari2017Coherent}, but these effects are beyond the scope of the current work.

Intuitively, intense laser pulses are expected to be more efficient in inducing enantioselective orientation. Indeed, for a single pulse excitation, a stronger pulse induces a higher degree of alignment (see Fig. \ref{fig:degree_alignment}), and one would think that it provides a basis for a higher degree of enantioselective orientation following the second pulse.
However, Fig. \ref{fig:figure1}(a) shows non-monotonic intensity dependence of the enantioselective orientation. 
Here, the optimal intensities are $I_1 = 16 \, \mathrm{TW/cm^2}$ and $I_2 = 37 \, \mathrm{TW/cm^2}$. At these intensities, $|\braket{\mu_Z}|_\mathrm{max}\approx 0.407\,\mathrm{Debye}$ (the degree of orientation is $\braket{\cos(\theta)}=\braket{\mu_Z}/\mu\simeq 0.199$).
At a given $I_2$, $|\braket{\mu_Z}|_\mathrm{max}$ increases with increasing $I_1$ when $I_1\le 16 \, \mathrm{TW/cm^2}$.
This trend is similar to the laser intensity dependence of $\braket{\cos^2(\theta_{\alpha_1 X})}_\mathrm{max}$ (see Fig. \ref{fig:degree_alignment}), which implies that $|\braket{\mu_Z}|_\mathrm{max}$ correlates with $\braket{\cos^2(\theta_{\alpha_1 X})}_\mathrm{max}$ to some extent.

\begin{figure*}[!t]
\centering{}
\includegraphics[width=\linewidth]{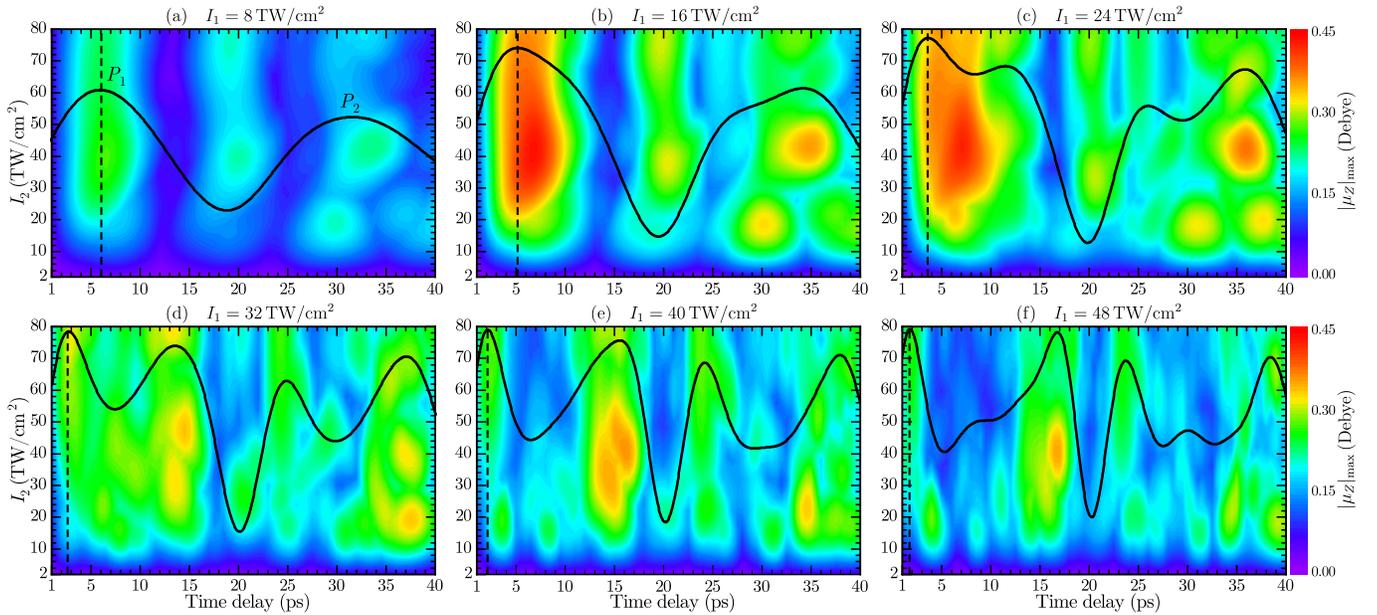}
\caption{Quantum mechanically calculated $|\braket{\mu_Z}|_\mathrm{max}$ as a function of $I_2$ and the time delay between the two laser pulses, $\tau$.
The peak intensity of the first laser pulse is fixed as (a) 8, (b) 16, (c) 24, (d) 32, (e) 40, and (f) $48\,\mathrm{TW/cm^2}$.
Here $T=0\,\mathrm{K}$ and $\sigma=50\, \mathrm{fs}$.
For comparison, the time-dependent degrees of alignment and the time of maximum degree of alignment  $\braket{\cos^2(\theta_{\alpha_1 X})}_\mathrm{max}$ for the case of Fig. \ref{fig:degree_alignment} are shown as solid  and dashed lines, respectively. The peaks of alignment are marked in (a). 
 \label{fig:2D_timedelay_I2}}
\end{figure*}

\begin{figure}[!b]
\centering{}
\includegraphics[width=8cm]{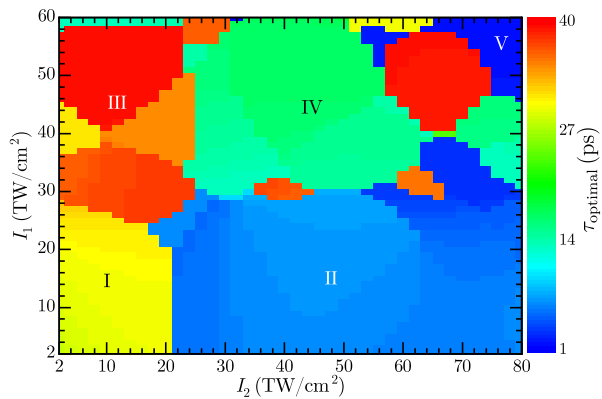}
\caption{Optimal time delay for the case shown in Fig. \ref{fig:figure1}(b) as a function of $I_1$ and $I_2$. 
\label{fig:optimal}}
\end{figure}

In the previous theoretical studies, high-intensity pulses were applied to molecules at non-zero rotational temperature, resulting in a relatively large number of populated rotational states and a rapid decay of the molecular alignment following the first pulse.  
Consequently, when chiral molecules were excited by the delayed cross-polarized laser pulses, the optimal delay appeared to be equal to the moment of maximal alignment. 
Here, in contrast, we consider the case of $T=0\,\mathrm{K}$ and relatively low-intensity pulses, which means that fewer rotational states are involved. Under such conditions, the degree of alignment has an approximately periodic behavior with several comparable maxima.
This allows achieving a high degree of enantioselective orientation when the second pulse is applied at one of several delayed alignment maxima.
In the general case, it is not a priori clear that the first (or global) alignment maximum is the best time for applying the second pulse to optimize the enantioselective orientation.

Next, instead of applying the second pulse at the moment of the first alignment maximum, we optimize the time delay $\tau$ $(\le 40\,\mathrm{ps})$ for each pair of intensities $I_1,\;I_2$. Figure \ref{fig:figure1}(b) shows the maximum of the dipole signal at the optimal time delay. The intensity dependence of $|\braket{\mu_Z}|_\mathrm{max}$ at the optimal time delay is close to the case shown in Fig. \ref{fig:figure1}(a), but with a higher degree of orientation. 
Another pattern is seen for $I_1 >30 \, \mathrm{TW/cm^2}$ where the optimal time delay is completely different from the moment of $\braket{\cos^2(\theta_{\alpha_1 X})}_\mathrm{max}$.
For the cases considered here, the optimal intensities are $I_1 = 19 \, \mathrm{TW/cm^2}$ and $I_2 = 42 \, \mathrm{TW/cm^2}$, and the induced dipole signal is about 0.455 Debye (the degree of orientation is $\braket{\cos\theta}\simeq 0.223$). 
At these laser intensities, the ionization yield of PPO molecules is $\ll 1\%$ (see Appendix \ref{app:ionization-yield}, Fig. \ref{fig:Ion_Yield}).

To further explore the effect of time delay on the enantioselective orientation, we fix the peak intensity of the first laser pulse, $I_1$, and plot the maximal induced dipole signal, $|\braket{\mu_Z}|_\mathrm{max}$, as a function of $I_2$ and the time delay (see Fig. \ref{fig:2D_timedelay_I2}). 
The time-dependent alignment signal induced by the first pulse is shown by the solid black line in each panel. 
For relatively low values of $I_1$, the degree of alignment oscillates and has two peaks within the first 40\,ps, (local maxima, $P_1$ and $P_2$) [see the solid lines in Figs. \ref{fig:2D_timedelay_I2}(a) and \ref{fig:2D_timedelay_I2}(b)].
For higher values of $I_1$, additional local maxima emerge.
A strong dipole signal (color-coded red) appears when the second pulse is applied near an alignment peak.  
$|\braket{\mu_Z}|_\mathrm{max}$ remains approximately the same for a wide range of intensities of the first pulse, $16\,\mathrm{TW/cm^2} \leq I_1  \leq 24\,\mathrm{TW/cm^2}$ [see Figs. \ref{fig:2D_timedelay_I2}(b) and \ref{fig:2D_timedelay_I2}(c)]. In both panels, the global maximum, $|\braket{\mu_Z}|_\mathrm{max}$ appears at very close values of $I_2$ and $\tau$.

Figure \ref{fig:optimal} summarizes the preceding analysis and shows that
the optimal time delay, $\tau_\mathrm{optimal}$ depends on both $I_1$ and $I_2$.
Generally, the optimal time delay is shorter for a stronger second laser pulse. Thus, Fig. \ref{fig:optimal} defines five regions in which the optimal time delay changes slightly and appears around peaks of alignment. 
In region I ($I_1< 24\,\mathrm{TW/cm^2}$ and $I_2< 20\,\mathrm{TW/cm^2}$), the optimal time delay is about 30\,ps, close to the moment of the second peak of $\braket{\cos^2(\theta_{\alpha_1 X})}$ [$P_2$, see Fig. \ref{fig:2D_timedelay_I2}(a)].
In region II, the optimal time delay is about 6.5\,ps, a bit greater than the moment of $P_1$. 
The optimal time delay in regions III and IV is about 39\,ps and 16\,ps, respectively, which correspond to different peaks of $\braket{\cos^2(\theta_{\alpha_1 X})}$.
Moreover, for the high values of $I_1$ and $I_2$ (region V), the optimal time delay is close to the moment of $\braket{\cos^2(\theta_{\alpha_1 X})}_\mathrm{max}$.
Quantum simulations show that the maximum degree of alignment appears shortly after the first laser pulse ($t<6.5\,\mathrm{ps}$), while the optimal time delay may appear around the moment of another peak of $\braket{\cos^2(\theta_{\alpha_1 X})}$. In contrast, in the classical case, the optimal time delay is determined by the first laser pulse and always appears at the time of $\braket{\cos^2(\theta_{\alpha_1 X})}_\mathrm{max}$, see Fig. \ref{fig:2D_classical} for the classically calculated maximum dipole signal as a function of $I_2$ and $\tau$ at fixed $I_1$.

\begin{figure}[!b]
\centering{}
\includegraphics[width=\linewidth]{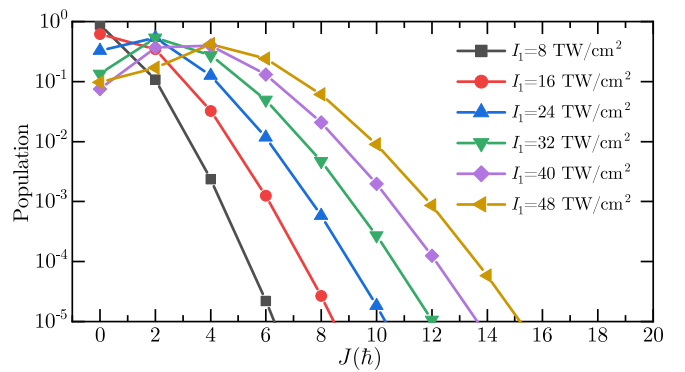}
\caption{Populations after the first laser pulse. Note that the populations of states with odd-$J$ are exactly zero.
 \label{fig:population_j_i1}}
\end{figure}

\section{Weak excitation limit}
Now, we focus on the enantioselective orientation in the deep quantum regime, when only few rotational states are involved in the dynamics.
At $T=0\,\mathrm{K}$, the molecules are initially in the ground rotational state ($\ket{JM\kappa}=\ket{0,0,0}$). The molecules are excited by non-resonant linearly polarized laser pulses.
Such pulses induce Raman-type transitions with $\Delta J=\pm 2$, thus only rotational states with even quantum number $J$ are excited.
Figure \ref{fig:population_j_i1} depicts distributions of populations as a function of quantum number $J$ after the first laser pulse, for several values of $I_1$. 
Higher laser intensities populate rotational states with higher $J$ number. For each laser intensity, the population decreases rapidly with increasing $J$ number. For example, at $I_1=48\,\mathrm{TW/cm^2}$, after the first laser pulse the first few states are significantly populated: 0.167 ($J=2$), 0.423 ($J=4$), and 0.242 ($J=6$); whereas for the low intensity $I_1=8\,\mathrm{TW/cm^2}$, only ($J=0$) is significantly populated.

Table \ref{tab:population2} summarizes the populations after the first laser pulse with $I_1=8\,\mathrm{TW/cm^2}$. Populations of states with higher $J$ number are negligible $(<0.5\%)$.
At this intensity, the majority of molecules remain in the ground state ($|0,0,0\rangle$).  
The time dependence of the molecular alignment is determined by the differences of energies of the rotational states excited by the first pulse. From Table \ref{tab:population2}, the dominant frequency component is $5.75\times 10^{-6}\,\mathrm{a.u.}$, or in terms of period $T_p=2\pi/(5.75\times 10^{-6})\,\mathrm{a.u.}$ which is about $26.4 \,\mathrm{ps}$.
Indeed, Fig. \ref{fig:2D_timedelay_I2}(a) shows that peaks $P_1$ and $P_2$ are separated in time by $\approx T_p$. This approximately periodic time-dependence allows achieving a high degree of enantioselective orientation by applying the second pulse at one of the local alignment maxima.
This is in sharp contrast to the classical case. Classically, the magnitude of alignment induced by a linearly polarized laser pulse decays rapidly, such that the optimal time delay is close to the moment of the maximum alignment emerging right after the first pulse (see Fig. \ref{fig:2D_classical}).

\begin{table}[!t]
\caption{Population (greater than 0.5\%) and energy of different rotational states after the excitation by the first laser pulse ($I_1=8\,\mathrm{TW/cm^2}$). 
Here $\kappa$ represents the renormalization of $K$, the projection of $J$ onto the molecular $z$ axis \cite{zare1988Angular}.
\label{tab:population2}}
\begin{centering}\setlength{\tabcolsep}{5mm}{
\begin{tabular}{ccc}\hline\hline
 \makecell[c]{States \\$|J M \kappa\rangle$}                & \makecell[c]{Population \\ (arb. units) }   
      &\makecell[c]{Energy \\($\times 10^{-6}$ a. u.)}                                \\ \hline
      $|0,0,0\rangle$   & 0.8896           & 0                  \\
      $|2,\pm 2,-2\rangle$  & 0.0291      & 5.75                 \\
      $|2,0,-2\rangle$   & 0.0194      & 5.75                 \\
      $|2,\pm 2,2\rangle$   & 0.0096      & 7.40                 \\
      $|2,0,2\rangle$   & 0.0064    & 7.40                \\
                   \hline\hline
\end{tabular}}
\end{centering}
\end{table}

\section{Finite temperature case}
\begin{figure}[!t]
\centering{}
\includegraphics[width=\linewidth]{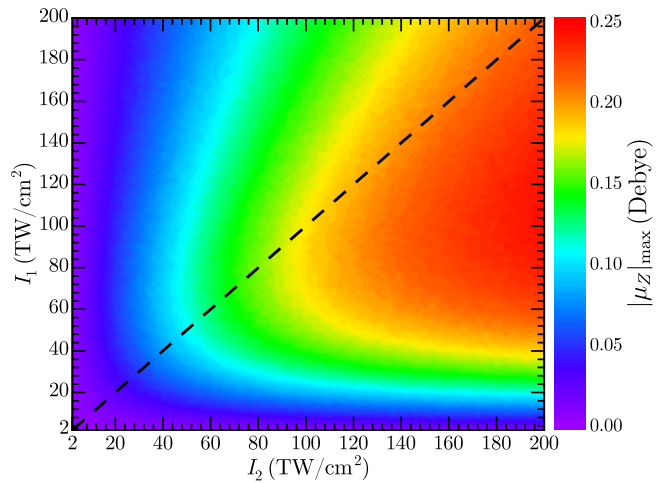}
\caption{Classically calculated 
$|\braket{\mu_Z}|_\mathrm{max}$ as a function of $I_1$ and $I_2$.
Here $T=5\,\mathrm{K}$, $\sigma=50\, \mathrm{fs}$, and the delayed laser pulse is applied at the moment of $\braket{\cos^2(\theta_{\alpha_1 X})}_\mathrm{max}$.  $N=10^6$ molecules are used to describe the molecular ensemble.
 \label{fig:temperature_5K}}
\end{figure}

In addition to the field parameters, the enantioselective orientation depends on temperature \cite{Tutunnikov2018Selective,Tutunnikov2019Laser}.
As an illustration, we consider the optimization problem at non-zero temperature. We use the classical treatment which happens to be in remarkable agreement with the quantum simulations for extended time intervals at elevated temperatures \cite{Tutunnikov2019Laser}.
Figure \ref{fig:temperature_5K} plots the classically calculated $|\braket{\mu_Z}|_\mathrm{max}$ at initial rotational temperature $T=5\,\mathrm{K}$.
Under these conditions, stronger laser pulses result in a higher dipole signal.
Compared with the case of $I_1=I_2$ (a diagonal dashed line in the figure), a stronger second pulse ($I_2>I_1$) is more effective for inducing a higher degree of enantioselective orientation, especially for the intensity $I_2$ greater than $100\,\mathrm{TW/cm^2}$ (the red region is below the line). This result means that for a given available total energy, more energy should be invested in the second pulse. 
At the given laser wavelength of 738\,nm and for the laser intensity lower than $160\,\mathrm{TW/cm^2}$, the ionization yield of PPO molecules is below 1\% (see Fig. \ref{fig:Ion_Yield}).
We find that at $T=5\,\mathrm{K}$, when $I_1=100\,\mathrm{TW/cm^2}$ and $I_2=160\,\mathrm{TW/cm^2}$, the induced dipole signal is about 0.227 Debye ($\braket{\cos\theta}\simeq 0.111$). 
In this case, the optimal time delay is $\tau=0.76\,\mathrm{ps}$.
Higher intensities can produce a higher degree of enantioselective orientation while inevitably leading to a non-negligible ionization yield.

\section{Conclusions}
In this work, we optimized the process of enantioselective orientation of chiral molecules excited by a pair of delayed cross-polarized femtosecond laser pulses. 
Our simulations show that at finite temperature (e.g., $T=5\,\mathrm{K}$), unequal strong laser pulses are needed to induce a sizable degree of enantioselective orientation.
At zero temperature, quantum effects lead to the existence of several optimal intensities maximizing the enantioselective orientation.
We also studied the dependence of the enantioselective orientation on the time delay between the two cross-polarized laser pulses. 
By optimizing the field parameters while maintaining the conditions for low ionization yield, we could achieve high degrees of enantioselective orientation at 0\,K ($\braket{\cos\theta}\simeq 0.223$) and at  5\,K ($\braket{\cos\theta}\simeq 0.111$).
It should be noted that the rotational energy transferred to the molecules depends on the total energy of the laser pulse. At the same time, a longer pulse with lower peak intensity will induce a lower ionization yield and may be preferable for enantioselectivity. The analysis of very long, picosecond laser pulses is beyond the scope of the current work.
A high degree of enantioselective orientation may be advantageous for enantiomeric excess analysis, as well as for enantioselective separation using inhomogeneous fields \citep{Yachmenev2019Field}. 
The enantioselective orientation along or against the propagation direction may be useful for further chiral discrimination through molecular ionization caused by  asymmetric fields, such as two-color laser pulses.

\begin{acknowledgments}
L.X. is a recipient of  the Sir Charles Clore Postdoctoral Fellowship. I.A. acknowledges support as the Patricia Elman Bildner Professorial Chair. This research was made possible in part by the
historic generosity of the Harold Perlman Family.
\end{acknowledgments}

\section*{Data Availability Statement}
The data that support the findings of this study are available from the corresponding author upon reasonable request.

\appendix

\section{Estimated ionization yield} \label{app:ionization-yield}
\begin{figure}[!h]
\centering{}
\includegraphics[width=\linewidth]{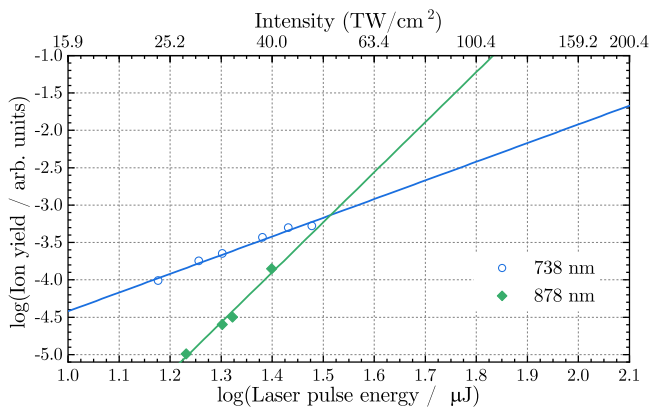}
\caption{Ionization yield of PPO molecules (and its linear regression) as functions of the laser pulse energy or the laser intensity. 
The marks represent the experimental data taken from \cite{Horsch2011Circular}.
 \label{fig:Ion_Yield}}
\end{figure}

Figure \ref{fig:Ion_Yield} depicts the ionization yield of PPO molecules driven by a circularly polarized femtosecond laser pulse. 
The intensity of the laser pulse is calculated for a pulse duration of 50\,fs \cite{Horsch2011Circular, Horsch2011Circular2}.
The ionization yield for a linearly polarized laser pulse is estimated to be of the same order of magnitude.
The figure shows that at $I=40\,\mathrm{TW/cm^2}$, the ionization yield is less than 0.1\% for the wavelengths of 738\,nm and 878\,nm.
The ionization yield grows with laser intensity.
The expected ionization yield at $I=160\,\mathrm{TW/cm^2}$ for 738 nm is about 1\%, see the blue line.
The low ionization yield allows us to neglect the ionization effects in our analysis.

\section{Classically calculated enantioselective orientation at zero temperature}

\begin{figure}[!b]
\centering{}
\includegraphics[width=\linewidth]{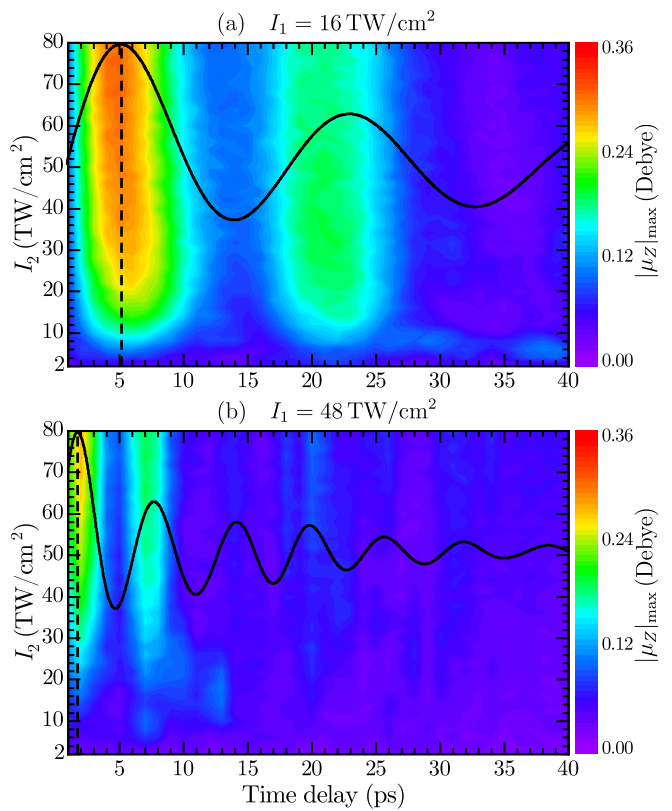}
\caption{Classically calculated $|\braket{\mu_Z}|_\mathrm{max}$ for the case shown in Figs. \ref{fig:2D_timedelay_I2}(b) and \ref{fig:2D_timedelay_I2}(f).
The peak intensity of the first laser pulse is fixed as (a) 16 and (b) $48\,\mathrm{TW/cm^2}$.
For comparison, the time-dependent degrees of alignment and the moment of $\braket{\cos^2(\theta_{\alpha_1 X})}_\mathrm{max}$ for $I_2=0$ are shown as solid  and dashed lines, respectively.
 \label{fig:2D_classical}}
\end{figure}

Figure \ref{fig:2D_classical} shows the classically calculated $|\braket{\mu_Z}|_\mathrm{max}$ as a function of $I_2$ and the time delay for the case shown in Fig. \ref{fig:2D_timedelay_I2} with $I_1=16,\, 48\,\mathrm{TW/cm^2}$.
A high degree of enantioselective orientation occurs when the time delay is close to the local alignment maxima induced by the first pulse. In contrast, the orientation vanishes when the time delay is close to alignment minima. 
In addition, the optimal time delay hardly changes with $I_2$ and appears at the moment of $\braket{\cos^2(\theta_{\alpha_1 X})}_\mathrm{max}$ since the magnitude of alignment decays rapidly. 
Moreover, the induced enantioselective dipole signal increases with $I_2$.

\bibliography{references}
\end{document}